\documentclass[english,12pt,preprint]{aastex}
\usepackage[]{natbib}
\usepackage{graphicx}

\shorttitle{Observations of 3C~66A with STACEE}
\shortauthors{Bramel et al.}

\begin{document}

\title{Observations of the BL Lac Object 3C~66A with STACEE}

\author{D.A. Bramel\altaffilmark{1}, J. Carson\altaffilmark{2}, C.E. Covault\altaffilmark{3},
P. Fortin\altaffilmark{4}, D.M. Gingrich\altaffilmark{5,6}, D.S.
Hanna\altaffilmark{4}, A. Jarvis\altaffilmark{2}, J. Kildea\altaffilmark{4},
T. Lindner\altaffilmark{4}, R. Mukherjee\altaffilmark{1,7}, C. Mueller\altaffilmark{4},
R.A. Ong\altaffilmark{2}, K. Ragan\altaffilmark{4}, R.A. Scalzo\altaffilmark{8},
D. A. Williams\altaffilmark{9}, J. Zweerink\altaffilmark{2}}

\altaffiltext{1}{Department of Physics, Columbia University, New York, NY, 10027}

\altaffiltext{2}{Department of Physics and Astronomy, University of California at Los Angeles, Los Angeles, CA 90095}

\altaffiltext{3}{Department of Physics, Case Western Reserve University, 10900 Euclid Ave., Cleveland, OH, 44106}

\altaffiltext{4}{Department of Physics, McGill University, 3600 University Street, Montreal, QC H3A 2T8, Canada}

\altaffiltext{5}{Centre for Subatomic Research, University of Alberta, Edmonton, AB T6G 2N5, Canada}

\altaffiltext{6}{TRIUMF, Vancouver, BC V6T 2A3, Canada}

\altaffiltext{7}{Department of Physics and Astronomy, Barnard College, New York, NY, 10027}

\altaffiltext{8}{present address: Lawrence Berkeley National Laboratory, 1 Cyclotron Road, Berkeley, CA 94720}

\altaffiltext{9}{Santa Cruz Institute for Particle Physics, University of California at Santa Cruz, 1156 High Street, Santa Cruz, CA 95064}

\begin{abstract}
We present the analysis and results of recent high-energy gamma-ray
observations of the BL Lac object 3C~66A conducted with the Solar
Tower Atmospheric Cherenkov Effect Experiment (STACEE). During the
2003-2004 observing season, STACEE extensively observed 3C~66A as
part of a multiwavelength campaign on the source. A total of 33.7
hours of data was taken on the source, plus an equivalent-duration
background observation. After cleaning the data set a total of 16.3
hours of live time remained, and a net on-source excess of 1134 events
was seen against a background of 231742 events. At a significance
of $2.2$ standard deviations this excess is insufficient to claim
a detection of 3C~66A, but is used to establish flux upper limits
for the source.
\end{abstract}


\section{Introduction}

To date, all confirmed extragalactic sources of TeV ($10^{12}$~eV)
photons have been low-redshift, high-frequency-peaked BL Lac objects
(HBLs) \citep{2004NewAR..48..527H}. Of all BL Lac objects, it is
reasonable that nearby HBLs, with their very energetic synchrotron
emission, would be the first to be detected at TeV energies. As the
energy thresholds of ground-based gamma-ray telescopes decrease, and
their sensitivities increase, there is significant potential for growth
in the number of very high-energy (VHE) gamma-ray sources to include
higher redshift objects and low-frequency BL Lacs (LBLs). The LBL
object 3C~66A is a likely candidate for detection with future VHE
telescopes because it was detected in the 30~MeV-20~GeV energy band
by the EGRET gamma-ray satellite instrument on the Compton Gamma Ray
Observatory, has a higher-energy synchrotron peak than most LBL objects,
and already has an unconfirmed TeV detection. 3C~66A also has a higher
redshift than any confirmed TeV source. In the context of the possible
absorption of gamma rays by intergalactic radiation fields, the greater
redshift of the source may indicate that detectors with improved sensitivity
to energies at or below 100~GeV, such as the Solar Tower Atmospheric
Effect Experiment (STACEE), will be able to confirm its detection.

3C~66A was first optically identified by \cite{1974ApJ...190L..97W}.
It is highly variable in the optical and X-ray bands \citep{1987A&A...178...21M}
and shows significant optical polarization \citep{1996A&AS..120..313T}.
It has been extensively observed in the radio and optical, but the
host galaxy surrounding the blazar jet has not been resolved. Observations
with the VLBA by \cite{2001ApJS..134..181J} show a highly superluminal
jet, confirming that beamed emission likely plays a major role in
the observed flux. The redshift for the source is widely quoted at
0.444, but the scant data supporting this value are highly uncertain
(see \S \ref{sec:Redshift}).

At MeV-GeV energies, 3C~66A is associated with the EGRET source 3EG
J0222+4253 \citep{1999ApJS..123...79H}. This association is not unique,
however, as the error box for 3EG~J0222+4253 also covers a nearby
pulsar, PSR~0218+42. The EGRET source position is consistent with
both 3C~66A and the pulsar, but the significance of the association
varies with energy \citep{2000A&A...359..615K}. Position contours
derived from low-energy (100-300~MeV) photons favor the pulsar location,
while high energy ($>1$~GeV) contours exclude the pulsar and correlate
well with 3C~66A. From this, it is concluded that the pulsar is the
primary source of the softer photons detected by EGRET, and 3C~66A
is the source of the harder component of the spectrum. Thus it can
be expected that the 3EG~J0222+4253 spectral index ($-2.01\pm0.14$) is a lower
limit for the 3C~66A spectral index, and that 3C~66A should produce
more high-energy photons than otherwise would be predicted.

Above the EGRET energy range, 3C~66A observations have been largely
unsuccessful. Repeated detections above 900~GeV by the Crimean Astrophysical
Observatory's GT-48 imaging atmospheric Cherenkov telescope \citep{1998AstL...24..134N,2002ARep...46..634S}
at an average integral flux level of $2.4\times10^{-11}$~cm$^{-2}$s$^{-1}$
have yet to be confirmed at other observatories. Observations by the
Smithsonian Astrophysical Observatory's Whipple 10-m gamma-ray instrument
produced a 99.9\% integral flux upper limit above 350~GeV of $<1.9\times10^{-11}$~cm$^{-2}$~s$^{-1}$
in 1993 \citep{1995ApJ...452..588K} and again of $<0.35\times10^{-11}$~cm$^{-2}$~s$^{-1}$
in 1995 \citep{2004ApJ...603...51H}, while observations by the HEGRA
telescope array in 1997 \citep{2000A&A...353..847A} produced a 99\%
upper limit above 630~GeV of $<1.4\times10^{-11}$~cm$^{-2}$~s$^{-1}$.
This uncertainty grants 3C~66A a `C-' rating in the TeV source summary
of \cite{2004NewAR..48..527H}, placing it among the most tenuous
of TeV sources.

As a potential gamma-ray source, 3C~66A is interesting for what it
could reveal about the density of extragalactic background light (EBL).
At TeV energies, gamma rays can interact with infrared EBL photons,
causing a decrease in the observed TeV flux that is related to the
column density of the EBL photons \citep{1999APh....11...93P,2001ApJ...555..641M}.
A redshift of 0.444 would place 3C~66A further away than any confirmed
TeV source to date and would make any observed TeV flux highly sensitive
to the effects of EBL absorption. Extrapolation of the 3EG spectrum
with no EBL absorption predicts a 100~GeV source flux around 0.2
Crab, but actual 3C~66A emission could in fact be higher given the
pulsar confusion in the 3EG source.

Modeling of the 3C~66A gamma-ray flux by \cite{2002A&A...384...56C}
predicts a moderate flux of $7.0-9.6\times10^{-11}$~cm$^{-2}$~s$^{-1}$
above 40~GeV, but negligible flux ($<0.14\times10^{-11}$~cm$^{-2}$~s$^{-1}$)
above 300~GeV. Without EBL absorption, 3C~66A has the potential
to be an easily detected TeV source and for this reason detections
and upper limits of the source are very useful for EBL studies. A
low predicted EBL cutoff energy of 100-200~GeV \citep{1999APh....11...93P}
puts 3C~66A out of reach for many older Cherenkov detectors, but
newer instruments may be able to push below the cutoff energy. 

The Solar Tower Atmospheric Cherenkov Effect Experiment (STACEE) is
a wavefront-sampling Cherenkov detector sensitive to photons above
100~GeV. No previous observations of 3C~66A have been reported by
any instrument with a gamma-ray energy threshold in the 100-300~GeV
range. The expected high-energy cutoff for 3C~66A is thus a unique
and interesting challenge for lower-energy ground-based Cherenkov
detectors such as STACEE.

The observations described in this paper were taken as part of a 3C~66A
multiwavelength campaign \citep{2004HEAD....8.0410B}. This campaign
took place during the 2003-2004 observing season and included optical
monitoring by the Whole Earth Blazar Telescope (WEBT) collaboration,
X-ray monitoring by the Rossi X-Ray Timing Explorer (RXTE), VHE gamma-ray
observations by STACEE and VERITAS \citep{2002APh....17..221W}, and
long-term radio monitoring. In addition, 9 high-spatial-resolution
observations using the VLA were carried out during the campaign and
throughout 2004 to follow possible structural changes of the source.
In this paper we describe only the high-energy gamma-ray observations
with STACEE (see also \citep{MyThesis}).

\section{The 3C~66A Redshift}

\label{sec:Redshift}The high redshift associated with 3C~66A is
one of the main features that sets it apart from most potential TeV
sources. The commonly quoted value of 0.444, we believe, may have
caused many in the TeV field to dismiss 3C~66A as an undetectable
source due to EBL considerations. As part of our work on this object,
we have conducted an extensive literature investigation of the widely
quoted redshift value and find that the data that back it up are remarkably
uncertain. Although not central to the results published in this paper,
it is important that the nature of this redshift value be made known
to the blazar community, particularly as 3C~66A may play a significant
role in future TeV measurements of EBL absorption.

To date, redshift measurements of 3C~66A have been reported in only
two papers. The first, \cite{1978bllo.conf..176M}, is most widely
credited with the z=0.444 measurement. Unfortunately, the redshift
value given in this paper for 3C~66A has been so widely quoted in
catalogs and summaries over the intervening years that the qualifications
of the original work have been generally disregarded. In the paper,
\citeauthor{1978bllo.conf..176M} call 3C~66A {}``one of two {[}objects{]}
in our study for which we feel we still have not measured a definitive
redshift'', and note of the single emission feature detected on the
source that {}``we are not certain of the reality of the feature.''
They summarize with the admonition that {}``the redshift 0.444 cannot
be considered reliable, and the object deserve{[}s{]} more attention.''

The second paper on the topic is that of \cite{1991ApJS...75..645K},
which reports on a reprocessing of International Ultraviolet Explorer
(IUE) data taken in the early 1980s. In this reanalysis it is noted
that a {}``weak feature near 1750 $\mbox{{\AA}}$ could be Ly$\alpha$
emission at the object redshift of 0.444'', but \citeauthor{1991ApJS...75..645K}
also indicate that \cite{1990PASP..102..463C} found a detector artifact
at the same location. For a point-source spectrum, \citeauthor{1990PASP..102..463C}
find that the region around 1750 $\mbox{{\AA}}$ shows an excess of
$2-3\times10^{-15}$~erg~cm$^{-2}$~s$^{-1}$~$\mbox{{\AA}}^{-1}$
that is created by the IUE instrument itself. The weak feature mentioned
by \citeauthor{1991ApJS...75..645K} is not more than $3\times10^{-15}$~erg~cm$^{-2}$~s$^{-1}$~$\mbox{{\AA}}^{-1}$
above the continuous source background. A close examination of the
3C~66A IUE spectrum in this region reveals a strong resemblance,
both in structure and in amplitude, to the IUE point-source artifact
spectrum. Without reanalysis that explicitly accounts for detector
artifacts, the ability of the IUE data to confirm the redshift is
questionable.

There are no other detections of emission lines for 3C~66A in the
literature, though there have been several attempts (e.g. \citet{1974ApJ...190L..97W,1976ApJS...31..143W}).
The data for both the initial detection and later confirmation of
the 0.444 redshift are not, in our opinion, sufficient to ensure a
reliable result. We find it reasonable to hold the value of 3C~66A's
redshift in question, and we encourage further spectroscopic observations
of this source.

\section{The STACEE Detector}

The Solar Tower Atmospheric Cherenkov Effect Experiment (STACEE) \citep{NIMpaper,ChantellEtAl}
is operational at the National Solar Thermal Test Facility near Albuquerque,
NM. It uses 64 large (37$m^{2}$) heliostat mirrors to focus Cherenkov
light from gamma-ray-initiated extensive air showers onto an array
of 64 photomultiplier tubes (PMTs), with each heliostat mapped onto
a unique PMT. High-speed electronics measure the charges and relative
arrival times of the PMT pulses. A multi-level coincidence trigger
\citep{MartinRagan} is used to select Cherenkov events. In this fashion
the detector samples the Cherenkov wavefront at 64 separate locations
on the ground, making STACEE a \emph{wavefront-sampling} detector.
Other gamma-ray detectors utilizing a similar technique include CELESTE
\citep{2002ApJ...566..343D} and Solar-Two \citep{2002AAS...200.2503T}.
The large mirror area obtained with the heliostats allows for the
detection of faint Cherenkov light pulses and grants a lower energy
threshold than all but the newest imaging-Cherenkov telescopes.

STACEE utilizes a two-level trigger to discriminate Cherenkov shower
events from randomly coincident night-sky background (NSB) photons.
The 64-channel array is broken up into eight clusters, each containing
eight PMT channels. The signals from each PMT are sampled by discriminators
set at a fixed threshold of approximately five photoelectrons. The
first-level (L1) trigger requires at least five out of eight channels
to have a discriminator hit within a 16~ns window. The second-level
(L2) trigger then requires at least five of the eight clusters to
trigger within the same 16 ns coincidence window before recording
a Cherenkov event. For each Cherenkov trigger, 1 GS/s Flash ADCs (FADCs)
are used to digitize the pulse on each channel, preserving all the
information contained in the wavefront sample.

STACEE has been fully
operational since 2001, and has detected gamma rays from the Crab
\citep{2001ApJ...547..949O} and Mrk~421 \citep{2002ApJ...579L...5B}.
It has also collected data on a number of active galactic nuclei \citep{JKHeidelberg}.
A full description of the STACEE detector can be found in \citet{GINGRICHNSS}.

\section{Data and Analysis}

Observations with STACEE are performed in on-off pairs, whereby a
source is observed for 28 minutes followed by a 28-minute observation
of an equivalent off-source area of dark sky. Each off-source observation
covers the same azimuth and elevation range as the corresponding on-source
observation. By using this technique, each instant of on-source data has
exactly one equivalent instant of off-source data, and comparisons between
halves of a pair are inherently corrected for observation-angle-dependent
systematics.  The difference between
the on-source and off-source shower-detection rates is attributed
to photons coming from the direction of the source. Since the off-source
observations are required for background estimation, an on-off pair
is considered to be the base unit of a STACEE observation. Background
for STACEE consists almost entirely of hadron-initiated air showers.

During the 2003-2004 observation season, 85 on-off pairs were taken
on 3C 66A. These pairs totaled 33.7 hours of on-source live time prior
to the data-cleaning process, plus an equal amount of off-source live
time. A series of data-cleaning criteria, or \emph{cuts}, were applied
to remove known hardware malfunctions, including removal of sections
in which one or more heliostats were not operational, the high voltage
system had tripped off, or parts of the data acquisition system were
inoperative. After the application of hardware cuts, a total of 29.3
hours on-source remained.

\subsection{Data-quality Cuts}

A set of cuts was applied to the data that were focused on removing
apparent or potential problems caused by unfavorable weather conditions.
A cut was made to remove all data that could potentially be contaminated
by frost build-up on the heliostat mirrors, as even a small amount
of condensation on the optics causes significant degradation in the
detector performance. The frost potential was determined by looking
at the records of the atmospheric frost index (defined to be the difference
between the atmospheric temperature and the dew point) during data
taking. Any data taken while the frost index was below $5.5\degr$
C were discarded.

A second data-quality cut was made to remove data biased by clouds
or changing atmospheric opacity. The L1 cluster rates are driven primarily
by accidental coincidences stemming from night-sky background (NSB)
photons and thus are sensitive to changes in background light levels.
At the STACEE site, light pollution is high enough that any increase
in sky opacity due to clouds, haze, or other atmospheric phenomena
causes a distinct increase in the NSB rate. Accordingly, the L1 trigger
rate provides a measure of the stability of the observing conditions.
Stable nights produce steady L1 rates, while unstable nights show
significant L1 rate fluctuations.

\begin{figure}
\epsscale{0.80}
\begin{center}\plotone{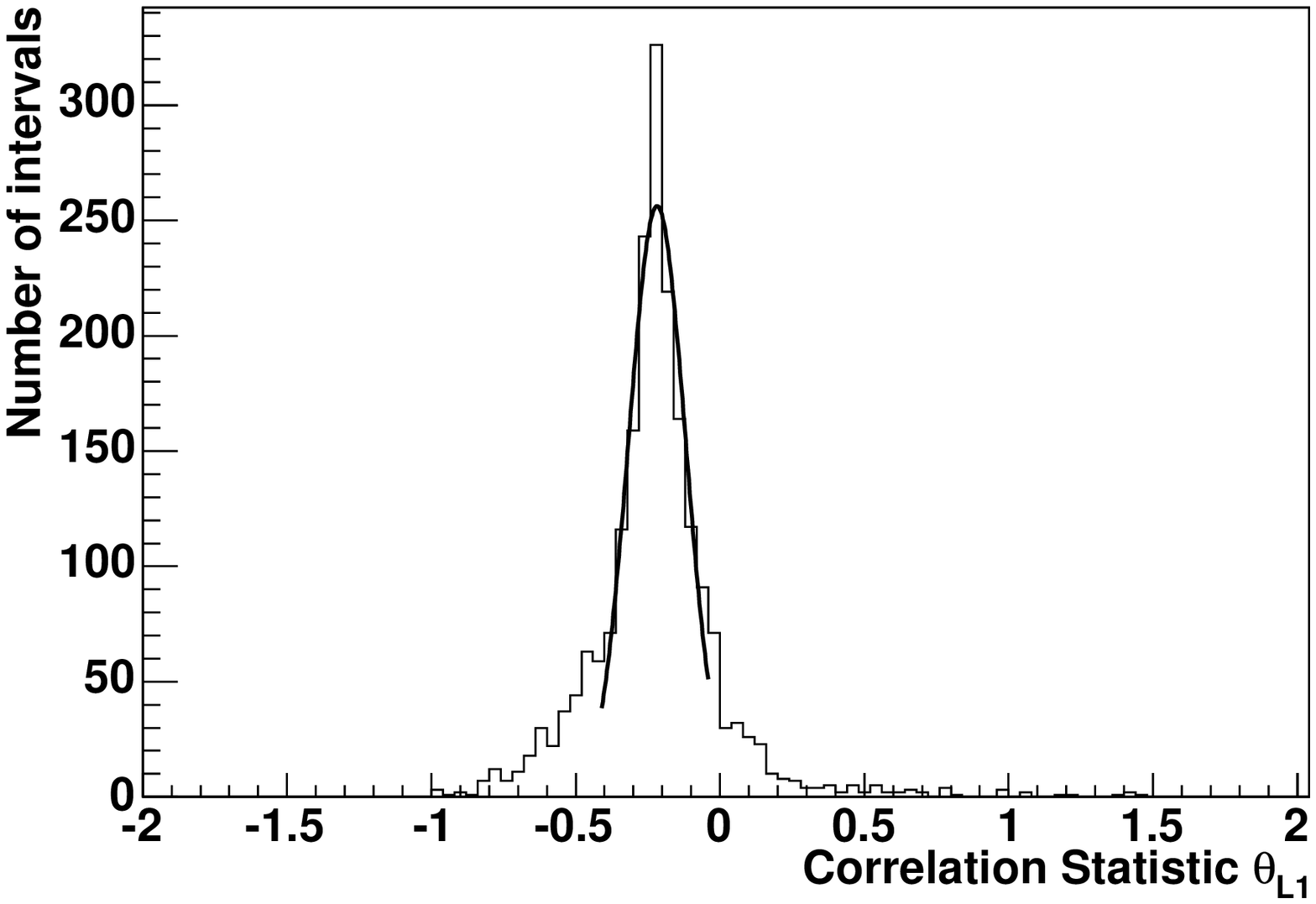}\end{center}

\begin{center}\plotone{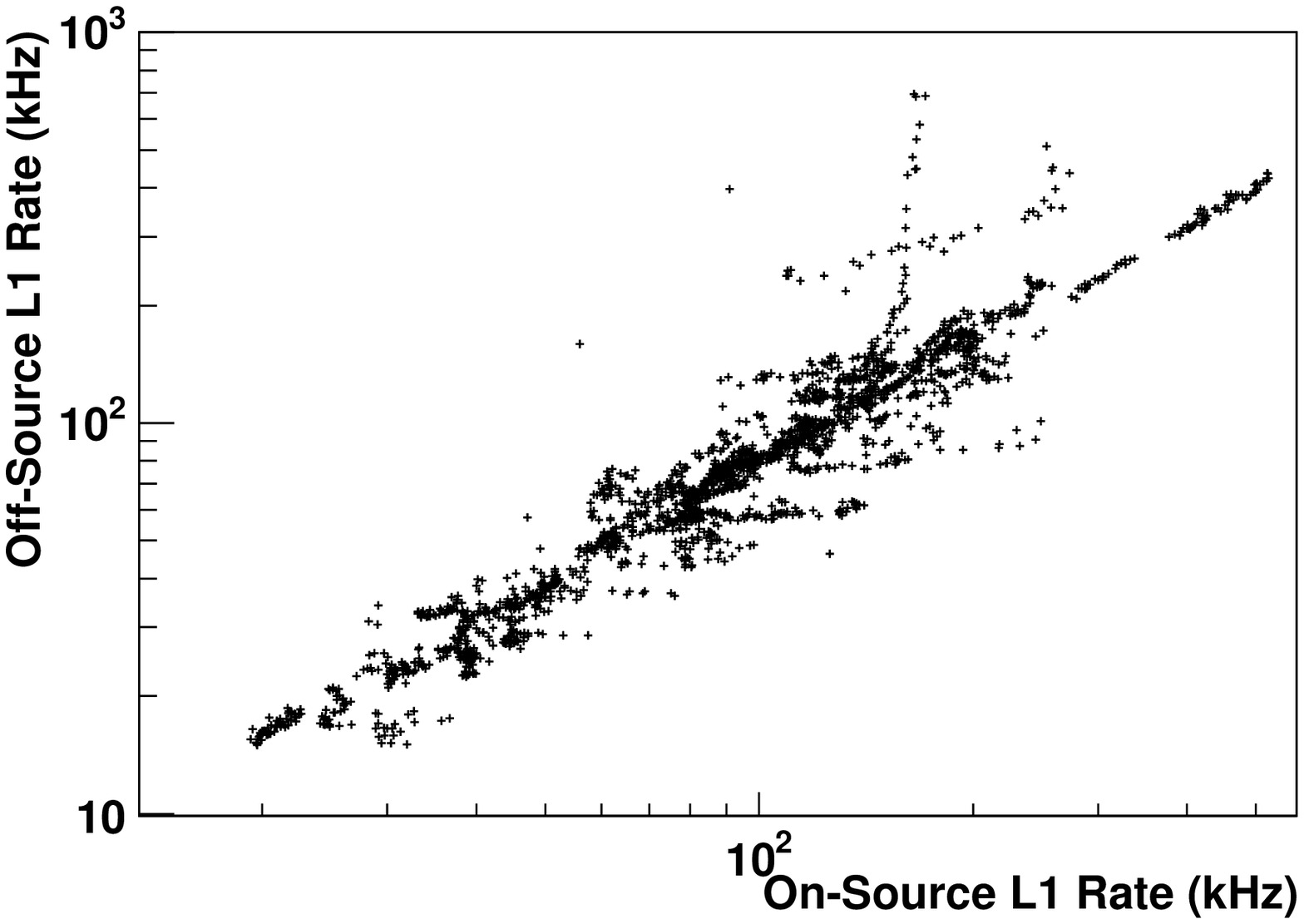}\end{center}

\caption{\label{fig: L1 Angle Graphs}Distribution of the L1 rate correlation
statistic for a single cluster over the entire 3C 66A data set. The
top panel shows a histogram of the correlation statistic. Normal variation
in the L1 trigger rates are shown by the central Gaussian peak of
the distribution, while unstable weather conditions cause the non-Gaussian
wings. The bottom panel shows a scatter plot of the L1 trigger rates
for on-source and off-source data. Changes in the ratio of the trigger
rates due to clouds and haze can be seen in the feathery fringes in
the scatter plot.}
\end{figure}

Since the NSB noise is elevation-dependent, no attempt was made to
constrain L1 rates to an absolute range. Instead, we used the cuts
to constrain the correlation between on-source and off-source L1 rates.
The data were divided into 30-second intervals, and the average L1
rates on- and off-source were calculated for each interval. A correlation
statistic $\theta_{L1}$ was defined for each interval based on the
L1 rates

\begin{equation}
\theta_{L1}=\ln\left(\frac{L1_{on}}{L1_{off}}\right)\label{eq: L1 Angle}\end{equation}
giving a characteristic measure of the correlation between the on-source
and off-source L1 rates. Histograms of this statistic for each cluster
were then compiled over the entire data set (see Figure \ref{fig: L1 Angle Graphs}).
For data that are unaffected by sky opacity changes, this distribution
of $\theta_{L1}$ for each cluster should be Gaussian, indicating
only random variations of the correlation statistic. In practice,
a Gaussian peak with extended tails is obtained, indicating non-random
variation caused by changing cloud cover. By performing a Gaussian
fit to the peaks of the distributions and discarding data in which
the correlation statistic of any cluster lies more than two standard
deviations from the mean value for that cluster, sections of data
that have been contaminated by clouds and haze were removed.%

A second, similar type of quality cut is made with the occupancy of
each channel. The occupancy of a channel is defined as the average
fraction of triggers in which the channel has a discriminator hit.
With stable observing and hardware conditions, the on-source and off-source
occupancies should be highly correlated. We define an occupancy correlation
statistic for a single channel similar to that of equation \ref{eq: L1 Angle}

\begin{equation}
\theta_{occ}=\ln\left(\frac{occ_{on}}{occ_{off}}\right)\label{eq: Occ angle}\end{equation}
and cut on the distribution of this quantity in the same way as with
the L1 statistic. Because of the larger number of channels that are
being constrained, there is a higher likelihood of a large fluctuation
in good-quality data than with the L1 cuts. To avoid eliminating these
data, we remove only data with three or more channels having occupancy
statistic values more than three standard deviations from the mean
value for that channel or with any one channel having an occupancy
statistic value more than five standard deviations from the mean value
for that channel.

After application of all data-quality selection cuts, a total of 16.8
hours of on-source data remained.

\subsection{Padding}

The measured background rate can be systematically biased by variations
in the brightness of the sky in the field of view of the detector,
often referred to as the \emph{field brightness}. When an observing
field is bright, the increased NSB photon count results in increased
levels of event promotions (dim Cherenkov showers that trigger simply
because they are boosted above threshold by random NSB photons). Thus
we expect bright fields to yield higher trigger rates than dark fields.

The field brightness can change between on-source and off-source observations
due to weather instability, but even very stable nights show a field
brightness difference due to the different stars in each field. The
brightness difference due to field stars is often rather small: the
number and magnitude of stars falling inside the $\sim1\degr$ STACEE
field of view tend to be roughly constant, and the promotion contributions
from each field cancel each other out. However, for certain sources
an individual bright star in one field upsets the balance (see, e.g.
\citet{2002ApJ...579L...5B}); the result is a highly distorted background
measurement that, if followed blindly, results in a spurious source
excess or deficit.

To counter the effects of field brightness differences, a technique
called \emph{software padding} is employed. Software padding effectively
increases the light level of the darker half of the pair by adding
a sample trace containing only NSB background to each event's FADC
trace. Once the light level is increased, a software trigger criterion is
applied to both halves of each pair. This padding technique, described 
in detail in \citet{2004ApJ...607..778S},
has been shown to effectively remove the effects of field brightness
differences in STACEE data. Data used in the 3C~66A analysis presented
here were padded using the waveform library technique of \citet{2004ApJ...607..778S}.
Cuts to remove sections of the data where the padding algorithm was
unable to execute were also applied, leaving a total net live time
of 16.3 hours on-source.

\section{Results}

\begin{figure}
\begin{center}\plotone{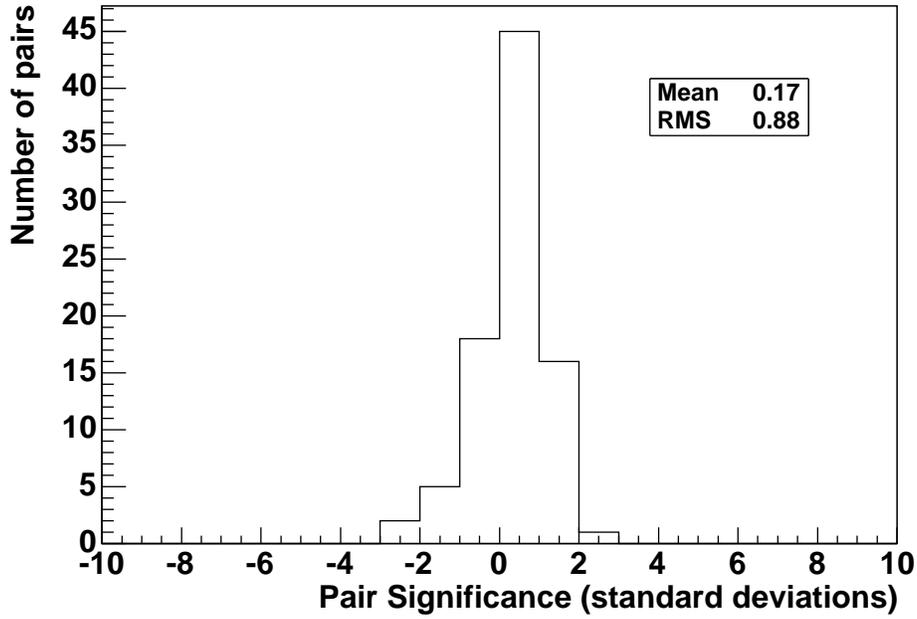}\end{center}

\caption{\label{fig: Pairwise Sigma}Histogram of the significance value,
in standard deviation units, for each on-off pair in the post-cut
data taken by STACEE on the source 3C 66A. The fact that there are
no on-off pairs having significance values far from zero indicates
that no significant transient events were seen.}
\end{figure}

After all time cuts and padding, we are left with 1134 excess on-source
events over a net observation time of 16.3 hours. The excess events
are against an off-source background of 231742 events, yielding an
on-source excess significance of $2.2$ standard deviations using
the method of \cite{1983ApJ...272..317L}. This excess significance
is not sufficient to claim a detection of the source. There were no
significant transient events in the data, as shown by the histogram
of significances for each of the 85 pairs in Figure \ref{fig: Pairwise Sigma}.%

\section{Detector Simulations}

In order to best understand the results of the observations of 3C~66A,
simulations of the STACEE detector were carried out to closely mimic
the 3C~66A data set. Using the CORSIKA air-shower simulation package
\citep{CORSIKA}, sets of showers were simulated with gamma-ray, proton,
and helium primaries. These were generated over a range of energies
and source hour angle ($\mathcal{H}$) positions. Cherenkov photons
from each shower were passed through a custom-made optics simulation
package and converted into photoelectrons by simulated PMTs. These
resultant photoelectrons were run through a custom-made electronics
simulation to determine which showers would have triggered the array.

Parameters for the custom simulation packages were specified to match
operational detector parameters as closely as possible. All fixed
detector parameters, such as coincidence conditions and electronics
configurations, were set in simulation to be identical to the detector
as it was during data taking. Variable parameters that affect detector
performance, such as PMT currents, were set to the average values
taken from data with hour angle similar to that of the simulated showers.
In this way, each simulated shower was processed with simulated parameters
as close as possible to the data it was intended to simulate. The 
simulation is able to reproduce several diagnostic quantities seen in 
the data, including L1 rates and cosmic-ray trigger rate as a function
of source hour angle.

Once the simulated showers were processed through the simulation pipeline,
analysis was carried out in the same manner as for real data.

\subsection{Effective Area}

To determine the sensitivity of the STACEE detector we have to determine
its effective area, the sensitive area presented to incoming gamma
rays. The effective area of the array was determined for each shower
type by scattering simulated showers across the detector, then multiplying
the triggered fraction by the total scattering area. Scattering areas
were circular in the plane normal to the arrival direction of the
primary particle, with radii sufficiently large that triggers near
the edge of the area were negligible.

Effective areas were calculated as a function of energy $E$ at several
hour angles $\mathcal{H}$ for each shower primary type. We interpolated
between the discrete effective area points and then weighted the effective
area by the source observation time to obtain a net effective area
as a function of energy\begin{equation}
A_{Eff}\left(E\right)=\frac{\int A_{Eff}\left(\mathcal{H},E\right)\times X\left(\mathcal{H}\right)d\mathcal{H}}{\int X\left(\mathcal{H}\right)d\mathcal{H}}\label{eq: Net Eff. Area}\end{equation}
where the exposure $X\left(\mathcal{H}\right)$ was taken from the
post-cut live time in the real data set. This net effective area,
computed separately for each primary particle type, is the effective
area of STACEE relevant for the 3C~66A data set.

\subsection{Energy Threshold and Flux Upper Limit}

\begin{figure}
\begin{center}\plotone{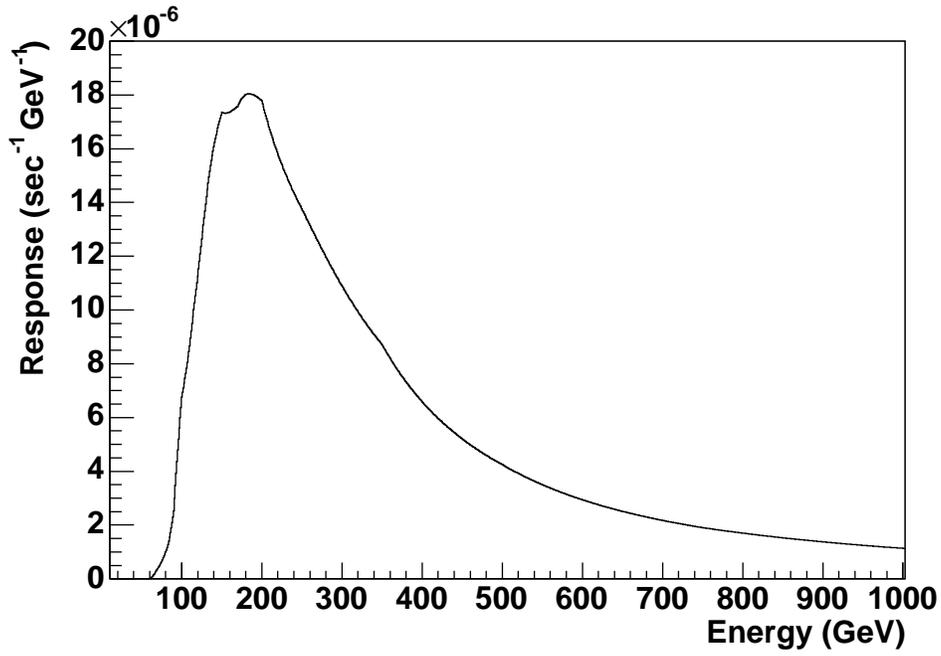}\end{center}

\caption{\label{fig: gamma response}Post-cut energy response curve of the
STACEE detector for an example power-law photon spectrum with a differential
spectral index of $-2.5$. The peak of the response curve defines
the energy threshold $E_{th}$ of the detector, though STACEE is sensitive
to photons below $E_{th}$.}
\end{figure}

Using the simulated effective area we calculated detector energy thresholds
and source flux upper limits. Following convention, the energy threshold
$E_{th}$ is defined as the peak of the response curve generated when
the effective area of the detector is convolved with a source spectrum,
as shown in Figure \ref{fig: gamma response}. The lack of any known
spectral properties for 3C~66A in the STACEE energy range adds a
degree of uncertainty to the estimation of the energy threshold and
flux upper limits. These estimates depend heavily on the assumed shape
of the source spectrum. At EGRET energies, the spectrum of 3C~66A
is known to be quite hard, fitting a power-law photon differential
index no softer than $-2.01\pm0.14$. At higher energies above 20 GeV the spectrum
no doubt falls off more steeply due to intrinsic softening and EBL
absorption, but neither of these effects is well constrained. The
amount of EBL absorption is particularly uncertain because the redshift
value of the source is, as argued above, very unreliable.%

We present in Table \ref{tab: Ethresh + Flux CI} the energy thresholds
and flux upper limits derived from the observations, for a variety
of power-law and EBL-absorbed power-law spectra.

\begin{deluxetable}{ccccc}

\tablewidth{5.0in}

\tablecaption{Integral Flux Upper Limits}

\tablehead{
      &\hskip 0.5in $\Gamma=\infty$ \tablenotemark{a}&   &  \hskip 0.5in
      $\Gamma=200$ \tablenotemark{a}& 
      \\
\colhead{Spectral Index}&\colhead{$E_{thresh}$\tablenotemark{b}}&\colhead{99\%CL\tablenotemark{c}}&\colhead{$E_{thresh}$\tablenotemark{b}}&\colhead{99\%CL\tablenotemark{c}}
}
\startdata

$-2.0$&$200$&$<1.0$&$150$&$<1.9$\\
       	                 
$-2.5$&$184$&$<1.2$&$150$&$<1.9$\\
       	                 
$-3.0$&$150$&$<1.7$&$142$&$<2.1$\\
       	                 
$-3.5$&$147$&$<1.8$&$137$&$<2.3$\\





\enddata

\tablecomments{99\% confidence limits on the 3C~66A photon flux derived in this work, assuming an EBL-absorbed power-law spectrum with photon index $\alpha$ and exponential EBL cutoff energy $\Gamma$.}

\tablenotetext{a}{Exponential EBL cutoff energy, in GeV.}

\tablenotetext{b}{STACEE energy threshold, in GeV.}

\tablenotetext{c}{Integral photon flux limit at the energy threshold, in units of $10^{-10}~cm^{-2}~s^{-1}$.}

\label{tab: Ethresh + Flux CI}

\end{deluxetable}

\section{Summary and Discussion}

The STACEE experiment observed the BL Lac object 3C~66A for a total of
33.7 hours in the fall of 2003. After all cuts and padding, 16.3 hours
of data yielded an on-source excess with a significance of 2.2
standard deviations, consistent with no detected flux. Flux upper
limits derived from simulated effective areas are given in Table
\ref{tab: Ethresh + Flux CI}.

\begin{figure}
\begin{center}\plotone{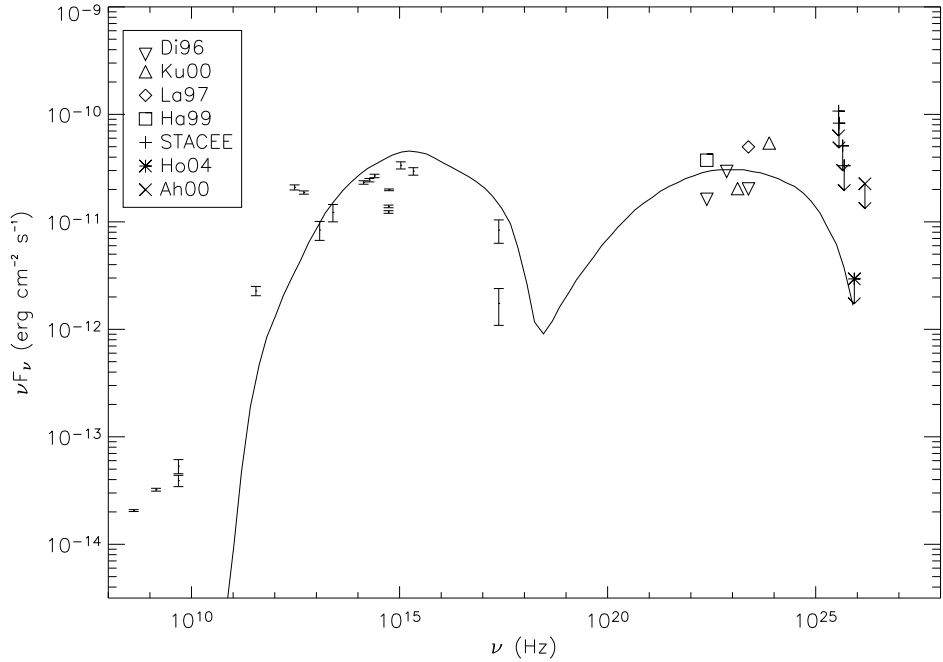}\end{center}

\caption{\label{fig: SED}The broadband spectral energy distribution
for 3C 66A. Radio - gamma-ray measurements have been compiled from the
literature. The EGRET detections are plotted as open symbols; Di96:
\cite{1996ApJ...467...589}, Ha99: \cite{1999ApJS..123...79H}, La97: \cite{1997ApJ...488...872}, Ku00:
\cite{2000A&A...359..615K}. The STACEE flux limits are shown as plus
signs. The TeV limits from Whipple and HEGRA are shown as stars and
crosses, respectively; Ho04: \cite{2004ApJ...603...51H}, Ah00:
\cite{2000A&A...353..847A}.}
\end{figure}

Figure \ref{fig: SED} shows the full spectral energy distribution for
3C~66A, made using non-contemporaneous data. 
Radio - X-ray data have been compiled from the literature and
are plotted with 1$\sigma$ error bars. Various EGRET detections are
shown as open symbols. These include the data of \cite{2000A&A...359..615K},
who estimated and subtracted the contribution from the nearby pulsar
PSR 0218+42; note that the
pulsar-subtracted measurement suggests that the high energy peak
continues to rise steeply into the STACEE energy band. 
The STACEE limits (plus signs) assume an unabsorbed spectrum with the
four spectral indices listed in Table 1. They are at a lower energy
threshold and higher flux than the TeV limits from the Whipple
telescope \citep{2004ApJ...603...51H} and HEGRA telescope array
\citep{2000A&A...353..847A}. A homogeneous, one-zone, synchrotron
self-Compton model from \cite{2002A&A...384...56C} is also shown. All
of the gamma-ray observations to date, including the STACEE flux
limits, are consistent with this model.

In conjunction with simultaneous data at other wavelengths, the STACEE
limits presented in Table \ref{tab: Ethresh + Flux CI} have the potential
to constrain models of the source emission mechanism. However, very
little physical modeling of 3C~66A has been published in the literature.
Modeling of high-energy gamma-ray emission from BL Lac objects is
very complex and depends extensively on simultaneous constraints from
lower-energy emission. The 2003-2004 multiwavelength campaign, of
which the STACEE observations are a part, will produce the first simultaneous
set of broadband spectral data on 3C~66A from optical to gamma-ray
energies.

Broadband modeling has been successful when applied to simultaneous
multiwavelength observations of sources similar to 3C~66A, such as
W Comae and BL Lacertae. All three of these BL Lacs share the similarity
that their synchrotron emission is dominant at X-ray energies. According
to the ROSAT all-sky survey data, the X-ray spectrum of 3C~66A is
soft (spectral index of 1.6) \citep{1998MNRAS.299..433F}, indicating
that the synchrotron component of 3C~66A extends to the X-ray regime.
The ROSAT measurement of a soft spectrum for 3C~66A is supported
by more recent observations by XMM \citep{2003MNRAS.346.1041C} and
by BEPPO/SAX \citep{2003A&A...407..453P}. Higher-energy synchrotron
emission by these LBLs indicates that they lie closer to the HBL end
of the BL Lac spectrum than most LBLs and thus are more likely to
have significant TeV emission.

In the past, W Comae and BL Lacertae have been the subject of
extensive multiwavelength campaigns. These data have been interpreted
using the fully time-dependent leptonic jet simulation code of
\citet{2002ApJ...581..127B} as well as hadronic Synchrotron-Proton
Blazar (SPB) models \citep{2002astro.ph..6164M}.  In the case of W
Comae, time-dependent modeling of the X-ray variability of the source
was found to yield quite different model predictions of the GeV-TeV
flux for hadronic or leptonic models \citep{2002ApJ...581..143B}.  W
Comae has thus proved to be a promising target for VHE gamma-ray and
coordinated broadband observations, as it may serve to distinguish
between leptonic and hadronic jet models for blazars. STACEE flux
limits on the $>$100~GeV emission from W Comae begin to constrain
hadronic emission models \citep{2004ApJ...607..778S}. Similarly, BL
Lacertae was also the subject of an extensive multiwavelength
monitoring campaign in 2000, and the data were modeled using leptonic
and hadronic jet models to fit the observed broadband spectra and
spectral variability patterns \citep{2004ApJ...609..576B}. Such
multiwavelength modeling studies with blazars tell us about cooling
timescales, magnetic fields, and Doppler factors associated with
blazar jets. Hadronic models for blazar jets predict significantly
greater TeV flux than leptonic models, and these differences can be
tested by future TeV observations of BL Lac by VERITAS or MAGIC
\citep{2004MmSAI..75..232M}.

Although such modeling is outside the scope of this current work,
studies are in progress to understand data from the recent 3C~66A
multiwavelength campaign using the fully time-dependent leptonic jet
simulation code of \citet{2002ApJ...581..127B} and to use the data
from radio through X-ray energies to make predictions about high-energy
emission \citep[in prep.]{3C_MW_model}. The implications of the STACEE
upper limits on 3C~66A could then be evaluated in the context of
the properties of the relativistic jet in 3C~66A, testing leptonic
and hadronic models for this source. Model-dependent flux predictions
at VERITAS energies can also be made, given a specific model and the
STACEE flux upper limits.

\cite{2003A&A...407..453P} argue that the multiwavelength spectrum and
modeling studies of 3C~66A could be applied toward resolving the
controversy regarding the identification of 3C~66A/PSR~0218+4232 with
the EGRET source. With the pulsar accounting for a majority of the
low-energy EGRET photons, the spectral energy distribution (SED) of
3C~66A in the EGRET energy range must be steeply
rising. \cite{2003A&A...407..453P} point out that it is difficult to
fit the 3C~66A SED from \cite{2000A&A...359..615K} to a log-parabolic
approximation of the synchrotron self-Compton (SSC) peak. A simple,
smooth model of the SSC peak favors a 3C~66A spectrum that included
the photons attributed to the pulsar by \cite{2000A&A...359..615K}.
Simultaneous modeling of the broadband 3C~66A spectrum with data at
X-ray and TeV energies would be able to help resolve the issue by
showing whether or not an additional component is able to fit the
\cite{2000A&A...359..615K} spectrum.

Detection of $>$100~GeV photons from 3C~66A would be a very interesting
result for gamma-ray astronomy. Currently all blazars detected by
TeV experiments have been high-frequency-peaked BL Lac (HBL) objects;
3C~66A is an LBL object, with a synchrotron peak at lower energies
than HBL objects. Confirmation of $>$100~GeV photons from this source
would open the door on a new class of objects for TeV astronomy. In
addition, if the redshift of 0.444 is correct, such a confirmation
would make 3C~66A the most distant known TeV source. This would make
it a very interesting source for constraining EBL spectral models.

It is our hope that the results presented here will motivate further
study of this source with a more sensitive detector. The possibility
that there may be detectable flux at energies near 100~GeV makes
3C~66A an interesting target for the new generation of imaging
Cherenkov telescopes such as VERITAS and MAGIC. Together with GLAST
\citep{2004HEAD....8.3006T} and AGILE \citep{2004HEAD....8.1605P} at
lower energies, observations of this source will provide a
comprehensive data set at sub-TeV gamma-ray energies that will be
important for modeling studies.

\acknowledgements{We are grateful to the staff at the National Solar Thermal Test Facility,
who continue to support our science with enthusiasm and professionalism.
This work was supported in part by the National Science Foundation,
NSERC (the Natural Sciences and Engineering Research Council of Canada),
FQRNT (Fonds Québécois de la Recherche sur la Nature et les Technologies),
the Research Corporation, and the University of California at Los
Angeles.}


\end{document}